\documentclass[pra,aps,showpacs,epsf,twocolumn]{revtex4}

\def\be{\begin{equation}}
\def\ee{\end{equation}}

\usepackage{graphicx}
\usepackage{bm}
\usepackage{amsmath}
\usepackage[T1]{fontenc}
\usepackage[latin9]{inputenc}
\usepackage{amsmath}
\usepackage{amssymb}
\usepackage{graphicx}
\usepackage{esint}
\usepackage[bookmarks=true,
   colorlinks=true,
   linkcolor=blue,
   urlcolor=blue,
   citecolor=blue,
   bookmarks=true,
   hyperindex=true
]{hyperref}
\allowdisplaybreaks[4]

\begin{document}
\title{Production of rubidium Bose-Einstein condensate in an optically-plugged magnetic quadrupole trap}

\author{Dong-Fang Zhang$^{1,3}$, Tian-You Gao$^{1,3}$, Ling-Ran Kong$^{1,3}$, Kai Li$^{1}$}

\author{Kai-Jun Jiang$^{1,2}$}
\email{kjjiang@wipm.ac.cn}

\affiliation{$^1$State Key Laboratory of Magnetic Resonance and Atomic and Molecular Physics, Wuhan Institute of Physics and Mathematics, Chinese Academy of Sciences, Wuhan, 430071, China\\$^2$Center for Cold Atom Physics,Chinese Academy of Sciences, Wuhan, 430071, China\\$^3$University of Chinese Academy of Sciences, Beijing 100049, China}

\date{\today}

\begin{abstract}
We have experimentally produced rubidium Bose-Einstein condensate in an optically-plugged magnetic quadrupole (OPQ) trap. A far blue-detuned focused laser beam with a wavelength of 532 nm is plugged in the center of the magnetic quadrupole trap to increase the number of trapped atoms and suppress the heating. A radio frequency (RF) evaporative cooling in the magneto-optical hybrid trap is applied to decrease the atom temperature into degeneracy. The atom number of the condensate is $1.2(0.4)\times10^5$ and the temperature is below 100 nK. We have also studied characteristic behaviors of the condensate, such as phase space density (PSD), condensate fraction and anisotropic expansion.
\end{abstract}

\maketitle

Since the experimental observation of Bose-Einstein condensate (BEC) in a dilute gas \cite{Wieman1995Science, Hulet1995PRL, Ketterle1995PRL}, the ultracold quantum gas has become a reachable tabletop to carry on a wide range of research, such as accurate measurement on physical constants \cite{Ye2014Nature, Ludlow2013Science}, precise spectroscopy \cite{Zelevinsky2015PRL}, quantum simulation \cite{Bloch2008RMP}, ultracold chemistry \cite{Jin2012ChemRev, Jin2011NaturePhysics}, equation of state of strong interacting quantum gases \cite{Ueda2010Science, Salomon2010Nature}, polaron behavior \cite{Zwierlein2009PRL, Salomon2009PRL}, spin-orbit coupling effect \cite{Spielman2011Nature, Jiang2012PRA}, and anisotropic character of the p-wave interaction \cite{Esslinger2005PRL, Jiang2014PRL}. Different kinds of external traps have been used to confine ultracold atoms where evaporative cooling is applied as a final stage to cool atoms into degeneracy, such as time-orbiting potential (TOP) magnetic trap \cite{Wieman1995Science}, quadrupole and Ioffe configuration (QUIC) magnetic trap \cite{Bloch1998PRA}, all optical trap \cite{Grimm2003Science, Chapman2001PRL}, and magneto-optical combination trap \cite{Spielman2009PRA}. Compared to many other kinds of traps, the optically-plugged magnetic quadrupole (OPQ) trap has many advantages, which has been demonstrated in some groups \cite{Ketterle1995PRL, Raman2005PRA, Perrin2012PRA, Shin2011PRA, Zwierlein2011PRA}. First, the tight confinement allows for fast radio frequency (RF) evaporative cooling and the large trapping volume offered by the magnetic quadrupole trap facilitates loading of a large number of atoms from the magneto-optical trap (MOT). Secondly, the atom cloud is positioned exactly in the center of the glass cell on the symmetry axis of the quadrupole coil pair, which ensures very good optical access to atoms. Third, the quadrupole coil pair allows to create large homogeneous magnetic fields by switching to copropagating currents, which we can use to address Feshbach resonances.

In this paper, we produce rubidium BEC in an OPQ trap. Here we show that the laser beam with a wavelength of 532 nm can be used to efficiently obtain $^{87}$Rb BEC despite the large detuning of the optical plug laser from the rubidium transition line at 780 nm. The atom number of the condensate is $1.2(0.4)\times10^5$ and the temperature is less than 100 nK, which matches the basic requirements for further advanced studies. We have also studied characteristic behaviors of the condensate, such as phase space density (PSD), condensate fraction and anisotropic expansion. In the near future we will use the obtained rubidium BEC to study the collective oscillations of the quantum gas, and generate an ultracold Bose-Femi mixture using the sympathetic cooling in the same experimental setup where we have cooled fermionic atoms like $^{6}$Li and $^{40}$K \cite{Jiang2013OL, Jiang2015PRA}.

\begin{figure}
\centerline{\includegraphics[width=0.93\columnwidth]{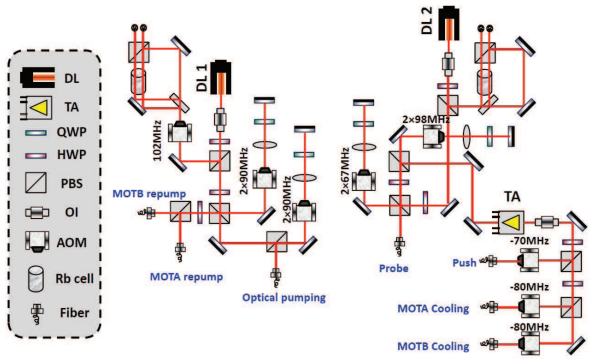}}
\caption{(color online) Optical arrangement for realizing BEC. DL: diode laser, TA: tapered amplifier, QWP: quater-wave plate, HWP: half-wave plate, PBS: polarization beam splitter, OI: optical isolator, AOM: acousto-optic modulator.}  \label{Fig1}
\end{figure}

We use the two-MOTs configuration to realize $^{87}$Rb BEC. The optical arrangement is shown in Fig.\ref{Fig1}. Two diode lasers (DLs) are individually frequency locked on the two crossover transitions $|F=1\rangle\rightarrow|F'=1,2\rangle$ and $|F=2\rangle\rightarrow|F'=2,3\rangle$, respectively, using the standard saturation absorption spectroscopy (SAS) method. DL2 affords cooling, probing and pushing beams, after passing through a series of acousto-optic modulators (AOMs) and being power amplified by one tapered amplifier (TA). The probing beam is resonant with the transition $|F=2\rangle\rightarrow|F'=3\rangle$ to probe the atomic population in the hyperfine state $|F=2\rangle$. DL1 provides repumping beams for MOTA and MOTB, respectively. The obtained BEC is in the spin state $|F=1,m_{F}=-1\rangle$, and the optical pumping beam pumps the atoms in hyperfine state $|F=1\rangle$ to the upper hyperfine state $|F=2\rangle$ for atom probing.

\begin{figure}
\centerline{\includegraphics[width=0.93\columnwidth]{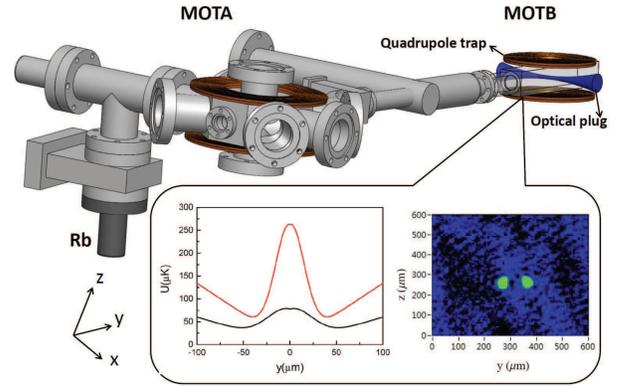}}
\caption{(color online) Schematics of the experimental setup. The whole system consists of two MOTs, MOTA and MOTB respectively. A rubidium oven is connected below the MOTA where a mechanical valve can switch on or off the atomic flux. In MOTB a pair of coils provide a magnetic quadrupole trap to combine cold atoms and a blue-detuned laser beam blocks the center of the trap to suppress the atomic spin-flip effect. Inset: the left plot shows the trapping potential of the OPQ trap along the $y$ axis for two sets of experimental parameters and the right plot indicates the {\it in-situ} image of two atomic clouds near the end of evaporative cooling. The red curve is for the plug beam power P = 5.0 W, the waist of the plug beam $\omega$ = 30 $\mu$m and the axial magnetic gradient $\partial B/\partial z = 200$ G/cm. The black curve is for P = 3.5 W, $\omega$ = 46 $\mu$m and $\partial B/\partial z = 90$ G/cm.}  \label{Fig2}
\end{figure}

The experimental setup is sketched in Fig.\ref{Fig2}. The main vacuum system consists of two chambers, the stainless chamber for MOTA and the glass chamber for MOTB, respectively. Two serial differential tubes with a diameter of 8 mm and a length of 100 mm are placed between the two MOTs, which pumps the vacuum pressure gradient with a magnitude of 2 orders. A rubidium oven (containing 4 g natural rubidium in which the abundance of $^{87}$Rb is 27.8\% ) is connected below the MOTA. A mechanical valve inserted between the rubidium oven and MOTA can switch off the atomic flux when experiment stops, protecting the ion pump from being polluted by absorbing too much rubidium from the vacuum background. The cooling process in MOTA uses three orthogonal laser beams retro-reflected in each direction for convenience, while MOTB uses six beams for easily controlling of the light-intensity balance. The number of cold atoms in MOTA is $4.0(2.1)\times10^8$ and the temperature is about 360(40) $\mu$K. The collected cold atoms in MOTA are pushed to MOTB with a pushing beam which is composed of a series of pulses with a period of 1.5 ms and a power of 1.0 mW.

The atom number in MOTB is $8.2(2.5)\times10^8$ and the temperature is about 320(35) $\mu$K. People generally use the compressed MOT and polarization gradient cooling (PGC) process to increase the atomic density and decrease the temperature, which will favor the magnetic loading \cite{Ketterle1995PRL, Bloch1998PRA}. Here we use the decompressed MOT \cite{Zwierlein2013PhD}, which can decrease the temperature and prevent atoms expanding too large simultaneously. In order to find the optimal experimental parameters we have studied the loading efficiency as functions of the magnetic field gradient, detuning of the cooling light and power of the repumping light. Decreasing the magnetic field gradient results in lower temperature, which increases the magnetic loading efficiency. But when the magnetic field is too small the atoms will expand fast, which decreases the magnetic loading efficiency. Finally we decrease the axial magnetic gradient from 18.0 G/cm to 8.4 G/cm, the power of the repumping laser from 10.0 mW to 0.3 mW, and increase the detuning of the cooling laser from -3$\Gamma$ ($\Gamma=2\pi\times6.1$ MHz is the natural linewidth) to -6$\Gamma$. The temperature is reduced to 130(20) $\mu$K which is sufficient for the magnetic loading. The atoms mainly stay in the lower hyperfine state $|F=1\rangle$ which is our target state for realizing BEC. After scanning the magnetic gradient to 336.0 G/cm with a period of 300 ms, we can load about $2.0(0.7)\times10^8$ atoms into the magnetic trap with a loading efficient of about 25\%. Only the spin state $|F=1,m_{F}=-1\rangle$ can be magnetically trapped and the temperature is 210(25) $\mu$K.

\begin{figure}
\centerline{\includegraphics[width=0.96\columnwidth]{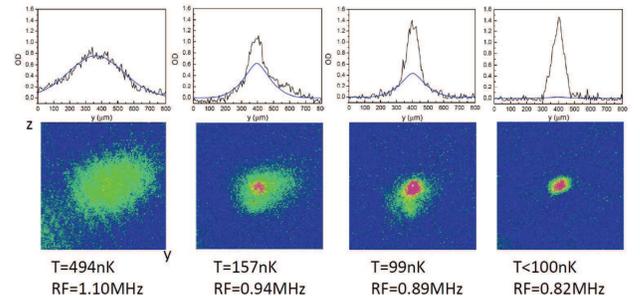}}
\caption{(color online) Producing BEC using the RF evaporative cooling in the OPQ trap. The upper row shows the line optical depth (OD) along the $y$ axis after a 20 ms time of flight (TOF). The blue curves are the Gaussian fitting to the two wings of the atomic density. The middle row shows the atomic density distribution in the $y-z$ plane. Different colors indicate the optical depth (OD) in the absorption imaging. The bottom row shows the RF frequencies and the corresponding atom temperatures.}  \label{Fig3}
\end{figure}

A far blue-detuned focused laser beam with a wavelength of 532 nm is plugged in the center of the magnetic trap to push away atoms from the zero point of the magnetic field, which suppresses the atomic spin-flip effect. The atomic lifetime in the trap is 71 s without the optical plug and 68 s with the plug. The similar lifetimes in these two cases indicate that the spontaneous-radiation heating due to the plug beam is negligibly small. As shown in the inset of Fig.\ref{Fig2}, we plot the trapping potential along the $y$ axis for two sets of experimental parameters. Two potential minimums obviously exist in the hybrid trap. Our calculation also indicates that the trapping frequency will increase when increasing the magnetic gradient or decreasing the waist of the plug beam, but the potential minimums will be closer to the center of the magnetic trap simultaneously. So we should make a compromise to provide a high enough trapping depth and suppress cold atoms from flying into the spin-flip area. Increasing the power of the plug beam can both increase the trapping frequency and the distance between the potential minimum and the zero point of the magnetic quadupole trap. In our experiment to produce BEC, the waist and the power of the plug beam are 30 $\mu$m and 5 W, respectively. Near the end of the RF evaporative cooling, the atom temperature is far below the optical barrier of the plug beam and the atomic cloud will spatially spit into two parts, which is shown in the inset of the Fig.\ref{Fig2}.

We shine a resonant probe beam on the atomic cloud along the x axis and monitor the atomic absorption images using a digital CCD. The RF evaporative cooling process is divided into four stages and totally has a period of about 24 s. Fig.\ref{Fig3} shows our obtained results near the end of the RF cooling with a 20 ms time of flight (TOF). Two BECs in the OPQ trap will merge into one during the TOF \cite{Ketterle1995PRL, Raman2005PRA, Perrin2012PRA, Shin2011PRA, Zwierlein2011PRA}, and we couldn't observe the interference fringe due to the weak coherence between the two BECs in our case \cite{Ketterle1997Science}. When the RF frequency is higher than 1.00 MHz, the main atoms are the thermal component which can be well fitted with a Gaussian distribution. The atomic density begins to deviate from the Gaussian distribution with the RF frequencies being lowered step by step, which means that the thermal and BEC components co-exist in the atomic cloud. The atom temperature can be extracted from the Gaussian distributions of the thermal component in different TOF. The minimum temperature that we can distinguish using this method is about 100 nK. When the RF frequency reaches 0.82 MHz, the pure condensate forms and the thermal component is negligibly small. We can numerically fit the density distribution of the condensate with a Thomas-Fermi function. The atom number of the pure condensate is $1.2(0.4)\times10^5$.

\begin{figure}
\centerline{\includegraphics[width=0.90\columnwidth]{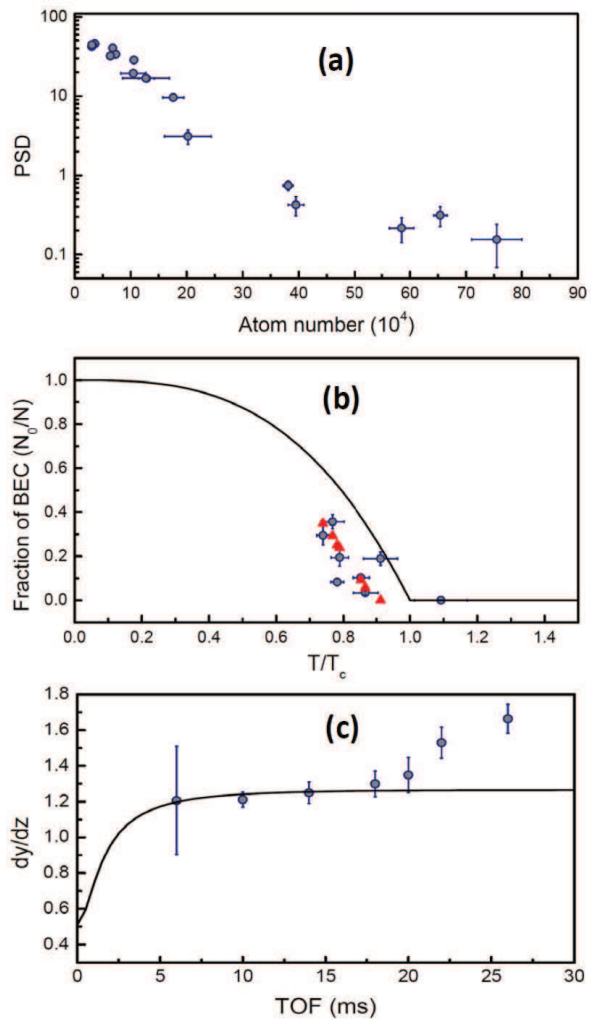}}
\caption{(color online) Characteristic behaviors of BEC. (a) The peak phase space density (PSD) versus the total atom number. (b) Fraction of condensate (N$_0$/N) versus the temperature normalized by the BEC threshold temperature T$_c$. The black solid curve is for the ideal gas: N$_0$/N = 1- (T/T$_c$)$^3$, the red triangles are for the numerical calculations of the interacting gases \cite{Raman2005PRA,Perrin2012PRA}, and the blue circles are for the experimental results. (c) Anisotropic expansion of the pure condensate in $y$ and $z$ axes in different TOF. The black curve is for the numerical calculation and the blue circles are for the experimental results. The error bars in all the plots indicate the standard deviation (SD)}  \label{Fig4}
\end{figure}

We have also studied characteristic behaviors of the condensate. In Fig.\ref{Fig4} (a), we show that the peak phase space density (PSD) increases with the total atom number decreasing during the RF evaporative cooling. The maximum value of the PSD we can achieve is about 44 which is much bigger than the BEC threshold value 2.612. Fig.\ref{Fig4} (b) shows the fraction of the condensate versus the temperature normalized by the BEC threshold temperature. The numerical calculation of the interacting gases \cite{Stringari1999RMP,StringariPRA1996} and the experimental results have a good agreement. Being limited by the measurement resolution, we only show experimental data with BEC fraction less than 40\%.  Fig.\ref{Fig4} (c) shows the anisotropic expansion in the $y-z$ plane in different TOF. Using the experimental parameters about the magnetic gradient and the plug beam, we can calculate the trapping frequencies in three directions: $\omega_x = 2\pi\times 60$ Hz, $\omega_y = 2\pi\times 233$ Hz, and $\omega_z = 2\pi\times 119$ Hz \cite{Ketterle1995PRL,Raman2005PRA,Perrin2012PRA}. The anisotropic expansion behavior is obvious in the experimental results. The main reason for the discrepancy between the experimental results and theoretical calculation is that the two atomic clouds will merge into one during expansion, while our calculation only considers one potential minimum in the trap. The accurate analysis about the anisotropic expansion in the OPQ trap needs further consideration. We will load the rubidium BEC from the OPQ trap into an optical trap which is composed of a far red-detuned laser beam with a wavelength of 1064 nm \cite{Raman2005PRA}. After turning off the OPQ trap and switching on the dipole trap simultaneously, the two BECs will merge into one because the dipole trap only has one potential minimum.

In conclusion, we have experimentally produced rubidium BEC in an OPQ trap. The laser beam with a power of 5 W and a wavelength of 532 nm can be used to efficiently obtain BEC despite the large detuning of the plug beam from the rubidium transition line at 780 nm. The atom number of the condensate is $1.2(0.4)\times10^5$ and the temperature is below 100 nK. We have also studied characteristic behaviors of the condensate, such as phase space density (PSD), condensate fraction and anisotropic expansion. This provides us the starting point to study the collective oscillations of the quantum gas and generate an ultracold Bose-Femi mixture using the sympathetic cooling. In addition, we can produce BEC with three spin states $|m_{F}=0,\pm1\rangle$ in the hyperfine state $|F=1\rangle$ by controlling the temporal sequence of switching of the magnetic trap and optical plug. We also can produce BEC using a 767 nm plug beam with a low power of about 100 mW. These results will be published elsewhere.

This work is supported by NSFC (Grants No. 11434015, No. 91336106, and No. 11004224) and NBRP-China (Grant No. 2011CB921601).

\end{document}